\documentstyle[aps,epsf]{revtex}

\begin{document}
\draft
\begin{twocolumn}
\title{Dynamical behavior of a dissipative particle in a periodic potential
subject to chaotic noise: Retrieval of chaotic determinism with
broken parity}
\author{Tsuyoshi Hondou
 \  and Yasuji Sawada}
\address{Research Institute of Electrical Communication \\ Tohoku University, 
   Sendai, 980-77 Japan}
\maketitle
\begin{abstract}
Dynamical behaviors of a dissipative particle in a periodic potential 
subject to chaotic noise are reported. 
We discovered a macroscopic symmetry breaking effect of chaotic noise on a dissipative 
particle in a multi-stable systems emerging,
 even when the 
noise has a uniform invariant density with parity symmetry and white Fourier 
spectrum. The broken parity symmetry of the multi-stable potential is not 
necessary for the dynamics with broken symmetry. We explain the mechanism of the symmetry 
breaking and estimate the average velocity of a particle under chaotic
noise in terms of unstable fixed points.
\end{abstract}

Success in explaining randomness of physical systems by 
the deterministic chaos is one of the most important progress of recent 
statistical physics.
 Tent map chaos is known to have the   
same randomness as the coin tosses which are universally presumed completely 
random~\cite{PhysicsToday}.
Therefore, it was assumed that 
the sequence of tent map chaos is far from its deterministic nature and  
that there is almost no difference between the chaotic sequence and 
the similar probabilistic random 
sequence except for their microscopic structure. 

 However, recent studies in complex systems indicate that there exists 
a condition in which undiscovered order of chaos emerges.
"Chaotic Itinerancy"~\cite{Arecchi,Otsuka,Kaneko} and  "Evolution to Edge 
of Chaos"~\cite{Kan,Bird} are some interesting examples.
The effect of chaotic dynamics on 
 information processing have also been studied
~\cite{Freeman,Aihara,Inoue,Nozawa,Maru}. 
It was also found that 
 the microscopic time correlation of chaos is important
 in learning process of chaotic time series by neural networks~\cite{Hondou}.
 One notices that these complex systems are generically multi-stable.
 Thus, it seemed that the effect of chaotic sequence 
on these multi-stable systems differs qualitatively from that of 
similar probabilistic random sequence.
 However,  high-dimensionality of the multi-stable systems of neural networks
 makes the study of the effect of chaotic sequence difficult.
Therefore it was required to select a simple multi-stable system of low
dimension in order to understand clearly the difference between the 
effects of chaotic sequence and probabilistic random sequence.
 In this paper, we demonstrate unexpected behavior of a dissipative particle 
in simple multi-stable systems subject to chaotic noise and clarify the 
reason for the peculiar behavior.

 Let us consider a particle  subject to an external chaotic stimulation 
in a periodic potential with an additional positive gradient as in Fig.1.
 Then a dissipative particle obeys the equation:
\begin{equation}
 \frac{dx}{dt}=-\frac{\partial V}{\partial x}+\sum_{j=-\infty}^{\infty} \xi_{j}
 \delta (t-j),
\label{eq:2}
\end{equation}
 where $\xi_{j}$ is a chaotic time series.
 The potential, $V(x) = V_{0}(x) +ax $, where $V_{0}$ is any periodic 
potential and $a$ is a constant. 
 In this paper, we report results of our study using a piecewise linear 
potential as a periodic potential $V_{0}(x) = h-(h/L) |x( \, {
\rm mod}(2L))-L| $ for $ x \ge 0 $, $ V_{0}(-x)=V_{0}(x) $, where  $L$ and $h$ 
are arbitrary constants; simply because comparison with a theoretical analysis
 is easy in this case. But it is easily verified that the central result is 
same for a smooth periodic potential. 
In the following we study a discretized equation,
\begin{equation}
  x_{n+1}=x_{n}-\frac{\partial V}{\partial x}|_{x=x_{n}}+\xi_{n} \, \, \, (n=0,
\, 1, \, 2, \ldots) 
\label{eq:3}
\end{equation}
which is obtained by integrating Eq.(1) from $t_{n}$ to $t_{n+1} = t_{n}+1$. 
It can also be verified that a choice of $\Delta t \equiv t_{n+1} - t_{n}$ 
does not essentially alter the following results.
In the present work, we mainly use chaotic time series produced by a tent map 
or a Bernoulli shift.
\begin{equation}
 \xi_{n} = - \eta_{n}.
 \label{eq:minus}
\end{equation}
\begin{equation}
 \eta_{n+1} =   f(\eta_{n}) = \left\{
\begin{array}{ll}
 -2|\eta_{n}|+1/2 &  (\mbox{Tent map}), \\
 2\eta_{n} -1/2  \, \mbox{sign}(\eta_{n}) & (\mbox{Bernoulli shift}).
\end{array}
\right.
 \label{eq:1}
\end{equation}
 The minus sign in Eq.(\ref{eq:minus}) is only for a demonstrative purpose.

It is easily verified that the invariant densities of the maps are constant, 
$\rho(x) = 1$ (for $-0.5 \le x < 0.5$, otherwise $\rho (x) = 0$), which is 
the same as the uniform random number $r_{n}$, $|r_{n}| < 0.5$; where
 invariant density, $\rho(x)$, is a solution of Frobenius-Perron integral 
equation: $\rho (y) =
\int dx \delta [ y - f(x) ] \rho (x) $~\cite{Schuster}.
  Therefore, in the limit of no potential barrier, 
the dissipative particle under either chaotic noise obeys Brownian motion 
without drift term; which is the same with the particle under the uniform 
random noise. In addition, the correlation function of the tent map is 
$\delta$-correlated which is also identical with the uniform random noise. 
 Thus it is useful to characterize the dynamical behavior of a system under 
chaotic noise by comparing with that under random noise~\cite{Gade}.

 One may naively guess that a particle under uniform random 
noise [-0.5:0.5] should move downward (left) in average, since noise is 
conventionally  used in multi-stable systems to realize global minimum state, 
with "annealing" as a typical technique~\cite{Kirk,Shinomoto}.
 But, surprisingly we discovered that the particle under chaotic noise produced 
by a tent map is 
driven upward (right) against the average potential gradient, which 
{\em never occurs} for probabilistic random noise (Fig.2).  

Hereafter we discuss for a while the case where parity symmetry of the
periodic potential holds; $a=\theta =0$.
Figure.~\ref{fig:nn} shows the numerically obtained transition probability 
of crossing  the potential barrier per unit time as a function of the  
potential width $2L$ for a fixed potential gradient  $h/L$.
The transition probability of a dissipative particle to 
cross the barrier is much higher under the action  of the tent map noise 
compared to the uniform random noise.
In the following we show that the hidden deterministic coherence of chaos
 is playing an important role to explain the observed phenomena.
 
One finds that a particle under the chaotic noise stays mostly in the 
neighborhood  of a basin of the potential. Thus the particle needs to 
be forced continuously by the noise having 
the coherent value to cross the barrier. This condition is satisfied when the 
chaotic noise
 stays in the neighborhood of an unstable fixed point. 
The nearer is the injected chaotic noise to the unstable fixed point, 
$\eta^{\ast}$,
 the longer $\eta$ stays in the neighborhood of $\eta^{\ast}$.
 By calculating the probability that the chaotic 
noise stays in the neighborhood of the unstable fixed point
 enough times 
to drive a particle to cross the barrier, one obtains the transition 
probability of the particle to cross the barrier as a function of the barrier 
width when the slope of the potential, $h/L$, is fixed and much smaller than 
the magnitude of the unstable fixed point: 

\begin{equation}
 P(L) \sim (1/\Lambda)^{L/|\eta^{\ast}|},
\label{eq:nn5}
\end{equation}
 where $\Lambda$ is the slope of the map ($2$ for tent map and Bernoulli 
shift)\cite{Journal}. 
 Numerical simulation for a tent map shows that $|\eta^{\ast}|$ is $0.4$,  
which should be contrasted with $0.5$ as the largest unstable fixed point for 
this map.
 The difference of the two values is attributed to the constant slope 
$h/L$ (= 0.1) which functions to decrease the climbing velocity of a particle.

 Equation.(5) is the general expression first derived for the barrier
crossing probability of a particle driven by a chaotic noise, characterized
 by the local Lyapunov index at the unstable fixed point. 
It explains why the symmetry break in the dynamics appears 
in the motion of the particle under a tent map noise. The value of 
$|\eta^{\ast}|$ for positive direction of the particle is $1/2$ and 
that of the negative direction is $1/6$. Thus the probability of crossing 
in the positive direction is much higher than the negative direction even 
when the potential has a parity symmetry $V(-x)=V(x)$. The fact that the 
transition probability in case of the tent map noise is much higher than 
in case of the uniform random noise can also be understood by the presence 
of the coherence of the noise in the former case while not in the latter case. 
Asymmetric motion against the average potential as shown in Fig.2 is a 
clear-cut result of the hidden order of the chaotic noise.

 A question arises how a dissipative particle in a periodic potential 
well behaves
under action of a chaotic noise produced by a map which has symmetric 
unstable fixed points. Numerical simulation showed that motion of a dissipative 
particle in the same potential under a sequence of noise created by a Bernoulli 
shift is not unidirectional but diffusive with a diffusion constant 
much greater 
than that under uniform random noise as shown in Fig.\ref{fi:dif}. The diffusive
 motion is explained by the equal transition probabilities to cross barriers 
in the positive and negative directions, as expected from the Eq.(~\ref{eq:nn5})
where the unstable fixed points are $\pm 1/2$ in the Bernoulli shift map. 
The difference
 of the diffusion constants between Bernoulli shift and uniform random  is 
again attributed to the presence of  
short-time correlation in the Bernoulli shift noise as can be seen from 
the difference 
of the transition probabilities of crossing the barrier between the cases of the
 Bernoulli shift and uniform random noise (see Fig.\ref{fig:nn}).

It should be noted, however, that the transition probability in case of 
Bernoulli
 shift is much greater than the case for the tent map (Fig.\ref{fig:nn}). 
Equation.(\ref{eq:nn5}) for 
the transition probability is applicable only when the mechanism for a particle
 to cross the barrier is dominated by a finite time series of monotonically 
increasing or decreasing chaotic noise. This is the case for the tent map 
but not 
for Bernoulli shift. In the latter case, there is a complex mechanism 
effective to increase the transition probabilities,
  in addition to the mechanism mentioned above. 
In case of Bernoulli shift, the chaotic noise 
starting from a neighborhood of an negative unstable fixed point can
 be reinjected 
to the neighbor of the same fixed point after one iteration of a small positive 
value as seen in Fig.\ref{fi:tentbe}. This chaotic series at least doubles the 
length of effective correlation and should be contrasted with the case of tent 
map, for which the time series must go through   a large  positive value to be  
reinjected into the neighborhood of the negative unstable fixed point. 
Therefore, 
the correlated motion of a particle climbing a slope would largely 
be interrupted.
 This observation qualitatively explains why the transition probability in 
case
 of Bernoulli shift is much larger than the case of tent map.

 Finally, to demonstrate the effect of the chaotic coherence more explicitly, 
we examined the effect of chaotic noise in the same potential discussed in 
Fig.~\ref{fig:2}. Here we used the noise produced by $\eta_{n+1} = f^{N}
(\eta_{n})$, where $f(\eta)$ is the tent map (ref. Eq.(~\ref{eq:1})) and $N$ 
is the time of iteration of the tent map. $N=1$ corresponds to the case
 discussed above.
Fig.~\ref{fig:4} shows the speed of the particle driven upward decreases
 as the number $N$ increases. 
This phenomenon is also explained by Eq.(\ref{eq:nn5}). The increase of $N$ 
corresponds to the increase of $\Lambda$ as $\Lambda = 2^{N}$. 
Thus, the 
increase of $N$ decreases transition probability as shown in Eq.(\ref{eq:nn5}).
 Thus, the systems having large Lyapunov exponent $\lambda$ ($\lambda =
 \log \Lambda 
= N \log 2$ in this case) lose the determinism of chaotic system; they 
approach to plain random systems. 
 This observation may be related to the recent computer simulation showing 
some chaotic system evolves to $\lambda = 0$ ("Edge of Chaos"), because the 
chaotic coherence work effectively in systems with relatively low Lyapunov 
exponent as demonstrated here.
 High performance of a Hopfield network with chaotic noise at the edge of 
chaos \cite{Maru} may also be explained by the present work.

 The drift/diffusion phenomena of chaotic systems without multi-stable
potential well have previously been 
discussed~\cite{forexam}.
It was reported in these literatures that difference of the drift/diffusion 
rate was caused by the differences of the invariant density or the initial 
condition of the chaotic noise. We demonstrated here that these quantities 
nor other ordinary statistical quantities such as correlation function of 
the chaotic noise do not explain the difference of the overall dynamics 
in multi-stable systems. Even if the higher order statistical quantities
are successful to demonstrate different characters of the noise, 
quantitative prediction or estimation 
of the dynamics
 under chaotic noise was found difficult by conventional 
approach\cite{RMP,ABC}.
However, the analysis of the effect of chaotic noise on multi-stable system
by focusing the unstable fixed points of the chaotic noise 
was found to be a new alternative tool to estimate such characteristics of the
dynamics.

 In this paper, we emphasized 
 that the emergence of symmetry breaking dynamics under chaotic noise is a special phenomenon 
in multi-stable systems. 
The multi-stability is widely observed in several fields of nature.
Protein motors are good examples.
 Magnasco showed the condition in which
 symmetry breaking dynamics (which models the motion of muscle)
 occurs when a dissipative particle moves in a periodic potential under 
correlated noise~\cite{Ratchet}. He proved that the
 asymmetry of periodic potential and time correlation of the driving 
force make it possible to produce broken symmetry dynamics.
 In contrast to his finding, we reported in this paper another condition for which broken 
symmetry dynamics emerges. 
 In this condition, asymmetry of the periodic 
potential is unnecessary if certain chaotic noise works as a driving force. The only condition to produce asymmetric motion
 is an asymmetric distribution of unstable fixed points of the chaotic noise. 
 The emergence of the broken symmetry dynamics under chaotic noise  
manifests itself as a typical example of the general findings that 
"multi-stability retrieves 
deterministic nature of chaos". 

 The authors would like to thank I.Nishikawa, T.Aoyagi, S.Sasa, H.Nozawa, T.Itayama, M.Yamamoto, M.Nakao, Y.Hayakawa and M.Sano for stimulating discussions.

\begin{figure}[h]
\caption{Periodic potential with constant gradient $a(=\tan \theta)$, 
period $2L$ and height $h$. One particle moves under this potential. 
In this figure $a=0.005$, $L=3.0$ and $h=0.3$.}
\label{fi:1}
\end{figure}

\begin{figure}[h]
\caption{Chaotic noise given by a tent map drives the particle upward 
against average potential gradient $a$, where $a=0.005$, $L=3.0$ and $h=0.3$. 
On the other hand, random noise drives the particle downward along the average 
potential gradient. Ordinate shows the distance scaled by the period of the 
potential. $x_{0} = 0 $ (initial value). The potential used is the same as 
in Fig.1.}
\label{fig:2}
\end{figure}

\begin{figure}[h]
\caption{
Probabilities $P_{tr}$ per unit time of a dissipative particle to cross 
a barrier under a tent map 
noise for various width $2L$ of potential barrier with a constant slope 
$h/L=0.1$ and no average gradient, $a=0$, obtained numerically.  
For comparative purpose, the transition
probability in cases of Bernoulli shift and random noise are also shown.
All noises have the same invariant densities and amplitudes.
 We also show a theoretical line given by Eq.(5) for the tent map.}
\label{fig:nn}
\end{figure}

\begin{figure}[h]
\caption{
Under action of noise generated by Bernoulli shift for the symmetric 
potential ($a=0$), the dynamics is a plain diffusion: $\langle x_{n}^{2} 
\rangle_{ensemble} \propto n$. But the diffusion constant under Bernoulli 
shift is much larger than that under uniform random [-0.5:0.5], where 
the two noises have the same invariant densities and amplitudes. 
Ordinate shows mean square distance scaled by the period of the potential, 
$2L$. Data subject to Bernoulli shift are ensemble averages  over $\eta_{0}$ 
(initial value of the map) of 100 systems, where $L=2$, $h=0.2$, $a=0$
 and $x_{0}$ = 0.
}
\label{fi:dif}
\end{figure}

\begin{figure}[h]
\caption{Typical sequence of (a) Tent map and (b) Bernoulli shift.
}
\label{fi:tentbe}
\end{figure}

\begin{figure}[h]
\caption{The trace of a particle motion versus time step. $N$ is the iteration 
number of the tent map. For $N \rightarrow \infty$,  the time series is  
entirely a random number. The potential and the initials used are the same 
as in Fig.2.
}
\label{fig:4}
\end{figure}
\end{twocolumn}
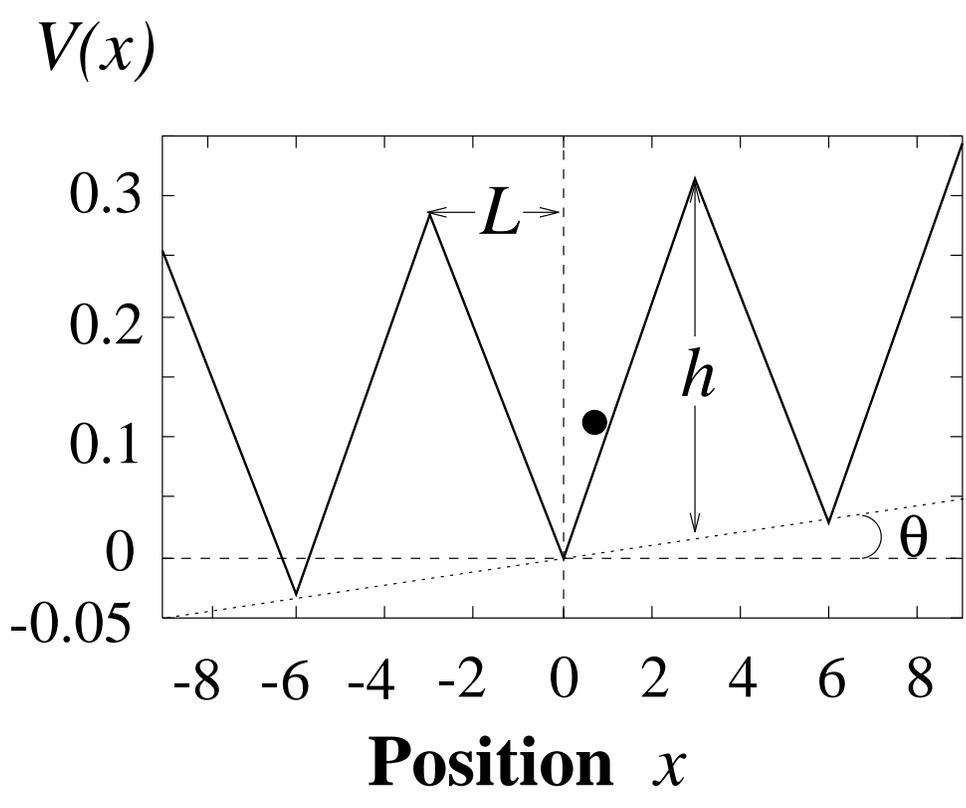
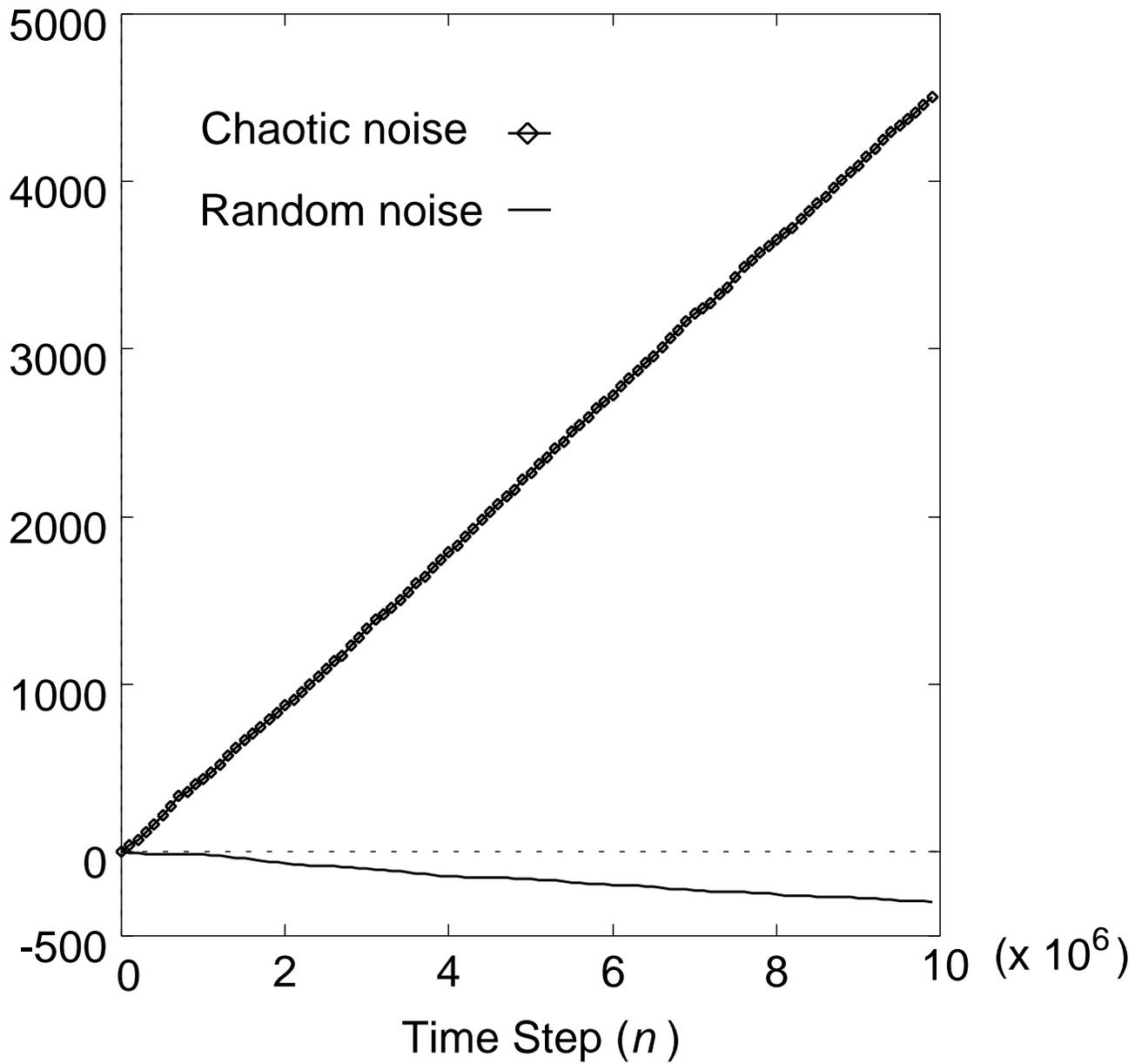
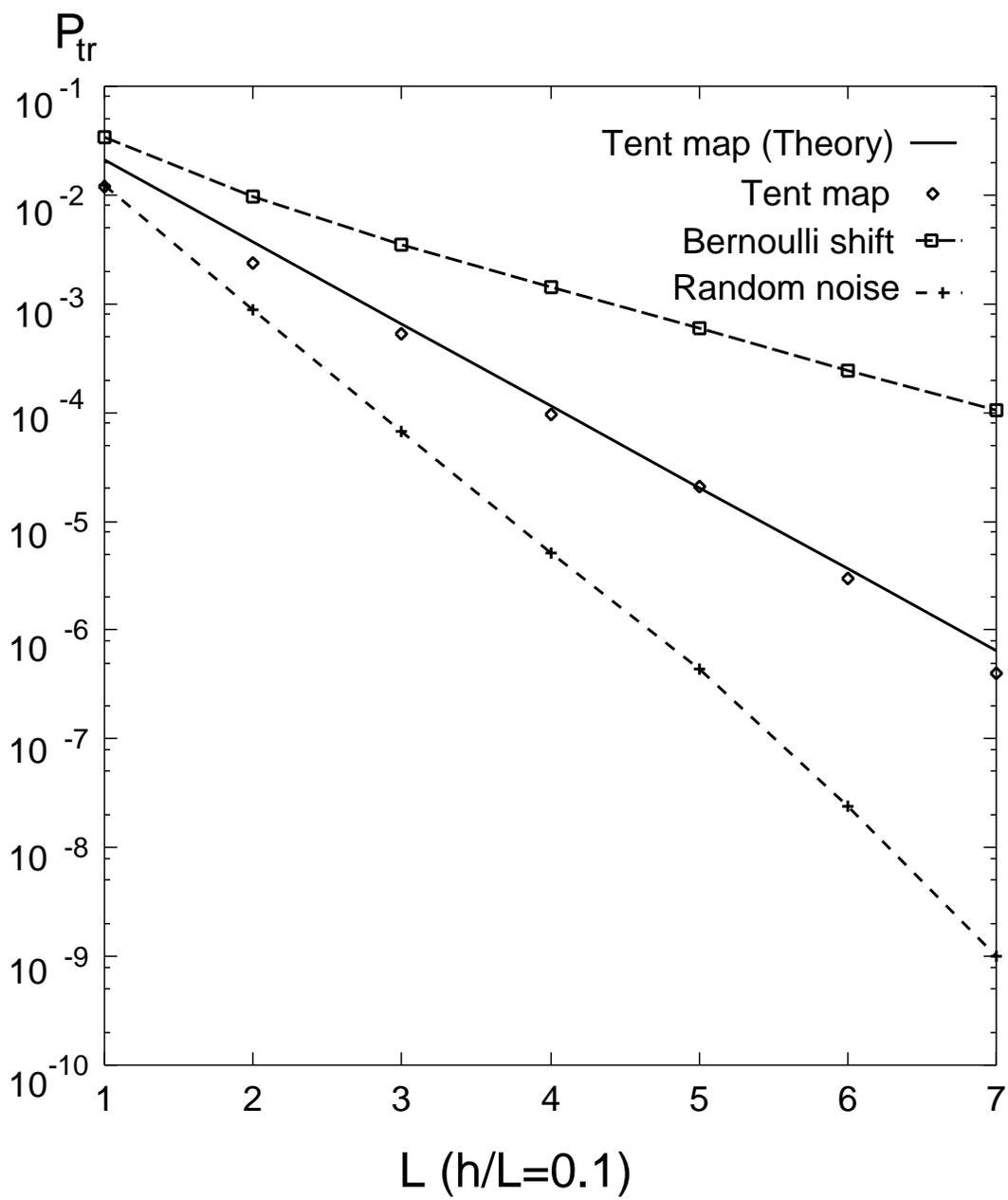
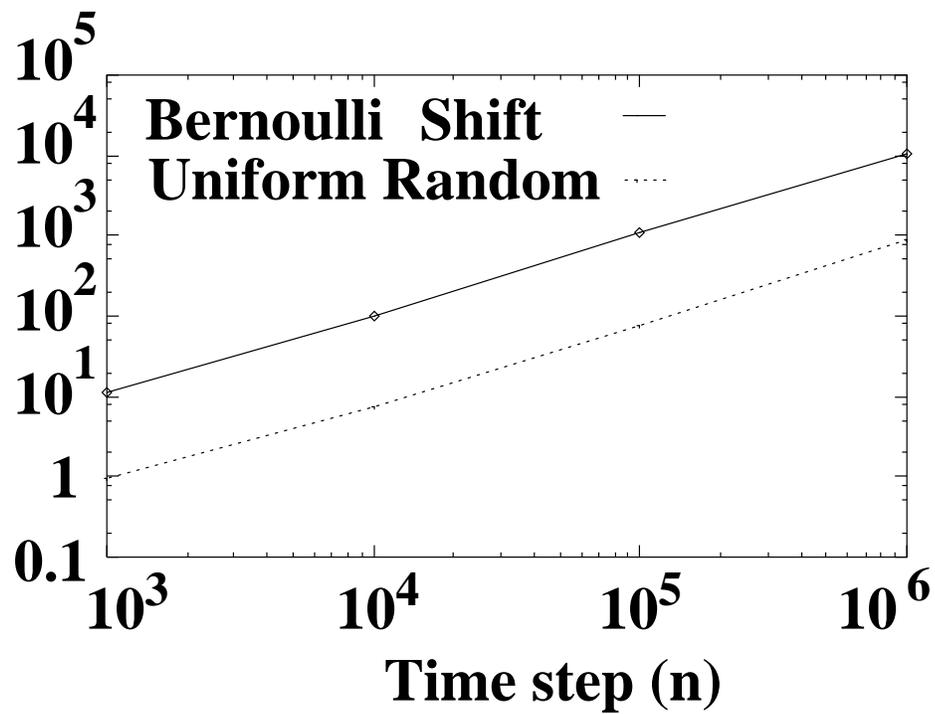
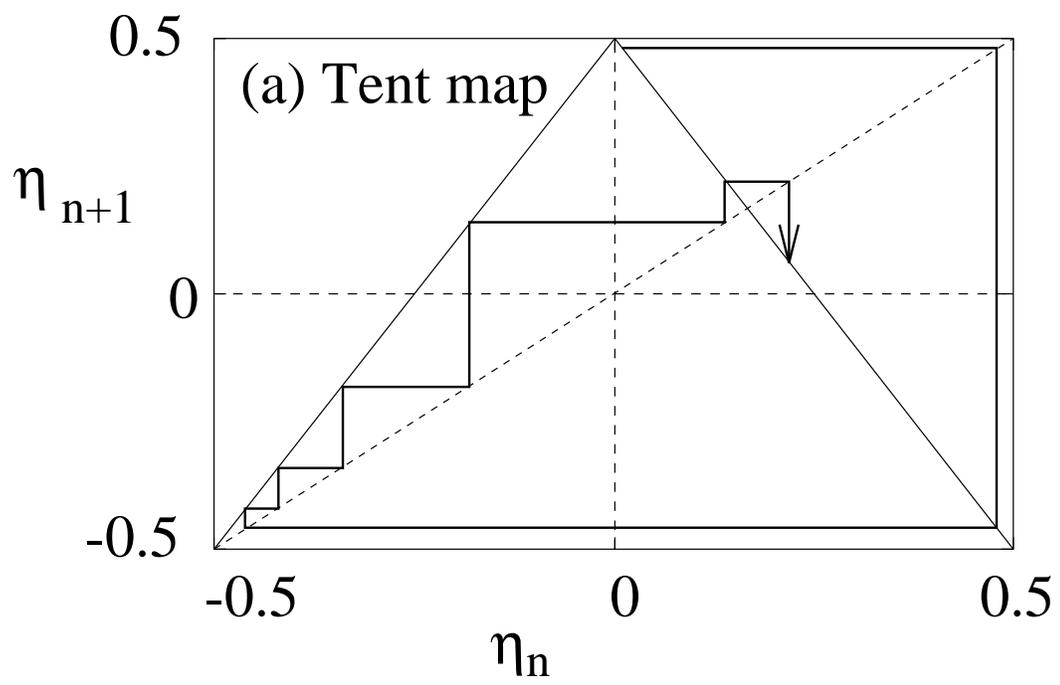
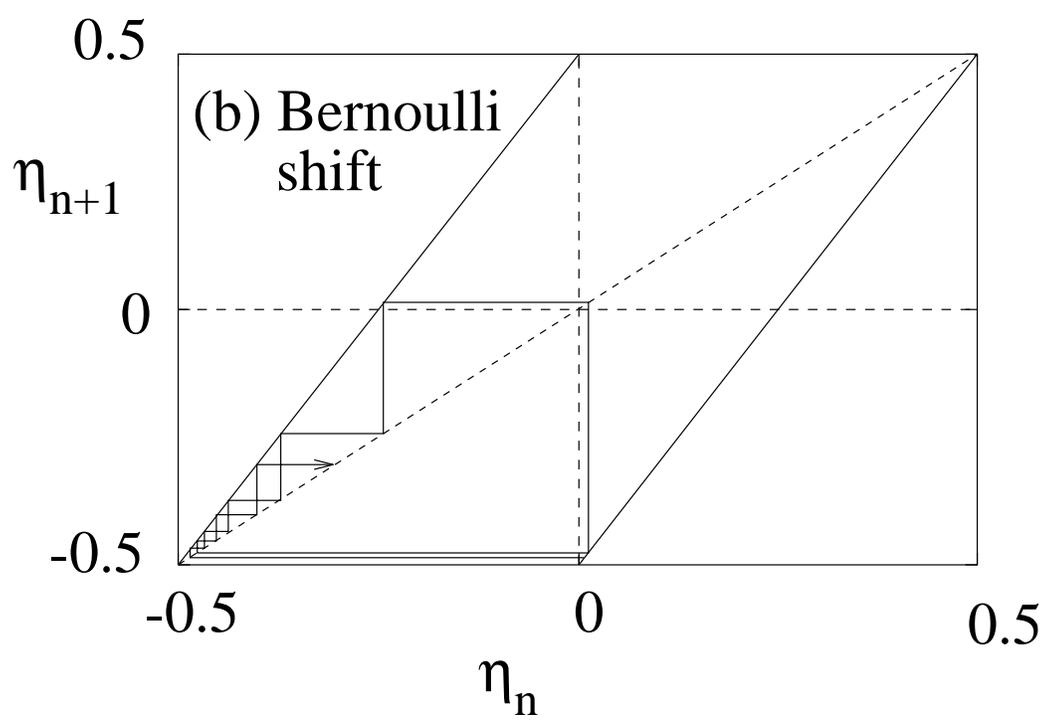
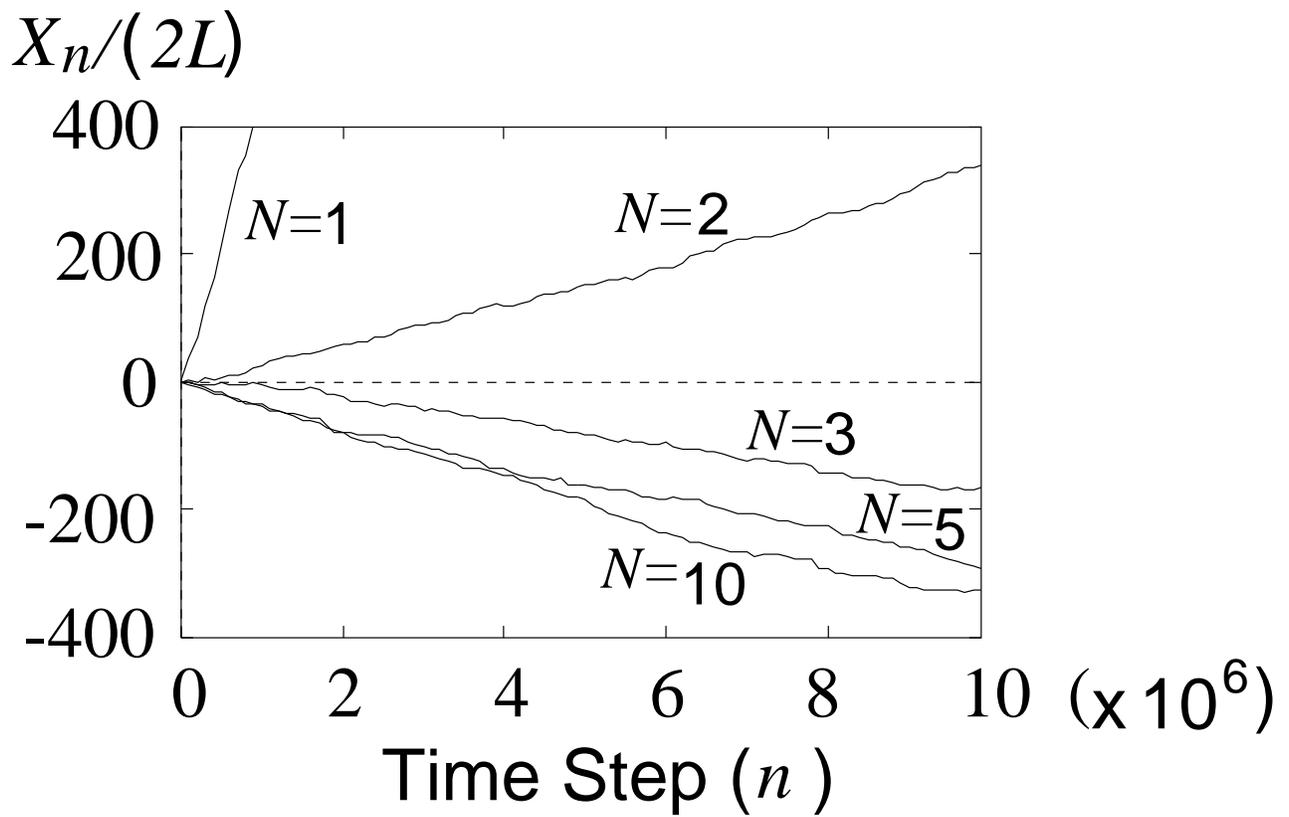
\end{document}